\documentclass[twocolumn,aps,showpacs,superscriptaddress,prl]{revtex4}
\usepackage{tipa}
\usepackage{bbding}
\usepackage{txfonts}
\usepackage{amssymb}
\usepackage{graphicx}
\usepackage{epstopdf}
\usepackage{epsfig}
\usepackage{float}

\begin{document}

\author{Liang Feng Huang}\affiliation{Key Laboratory of Materials Physics, Institute of Solid
State Physics, Chinese Academy of Sciences, Hefei 230031,
China}\affiliation{Department of Materials Science and Engineering, Northwestern University, Evanston, Illinois, 60208, USA}
\author{Peng Lai Gong}\affiliation{Key Laboratory of Materials Physics, Institute of Solid
State Physics, Chinese Academy of Sciences, Hefei 230031, China}
\author{Zhi Zeng}
\email{zzeng@theory.issp.ac.cn}\affiliation{Key Laboratory of
Materials Physics, Institute of Solid State Physics, Chinese Academy
of Sciences, Hefei 230031, China}\affiliation{University of Science
and Technology of China, Hefei, 230026, China}

\title{Phonon Properties, Thermal Expansion, and Thermomechanics of Silicene and Germanene}

\begin{abstract}
We report a hierarchical first-principles investigation on the
entangled effects of lattice dimensionality and bond characteristics
in the lattice dynamics of silicene and germanene. It is found that
bond bending (stretching) negatively (positively) contributes to
Gr\"uneisen constant $\gamma$, which results in the negative
acoustic (positive optical) $\gamma$. The layer thickening (bond
weakening) caused by chemical functionalization tends to increase
(decrease) the acoustic (optical) $\gamma$, due to the increased
(decreased) bond-stretching effect. The excitation of the
negative-$\gamma$ modes results in negative thermal expansion, while
mode excitation and thermal expansion compete with each other in
thermomechanics. The sensitive structural and electronic responses
of silicene and germanene to functionalization help us to derive a
generic physical picture for two-dimensional lattice dynamics.
\end{abstract}


\maketitle

\par Two-dimensional silicon and germanium (silicene and germanene)
\cite{Takeda50} have attracted great interest after the advent of
graphene \cite{Novoselov306}. The low dimensionality of graphene
\cite{Geim83} renders its charge density, electronic states, and
various physical and chemical properties readily controllable
\cite{Geim83,Singh56}, which is highly helpful for advanced
materials and devices. Apart from having these low-dimensional
benefits, silicene (Si) and germanene (Ge) are also expected to be easily
incorporated into existing silicon-based industry
\cite{Xu113,Jose47}.

\par Si and Ge have buckled hexagonal lattices \cite{Cahangirov102,Sahin80},
where quantum Hall effects, valley polarization, tunable band gap,
and fast intrinsic mobility have been discovered
\cite{Liu107,Ezawa109,Tsai4,Drummond85,Ni12,Pan112,Li2013}. This
indicates their promising applications in electronics, optics, and
fundamental-physics research. Si has been synthesized on various
substrates (Ag \cite{Lalmi97,Vogt108,Kara67}, Ir \cite{Meng13}, and
ZrB$_2$ \cite{Fleurence108}), while Ge is still under exploration.
Monolayer Si transistors operating at room temperature have been
fabricated recently \cite{Tao2015}. Si and Ge can be easily
functionalized in various environments. Chemical functionalizations
(e.g., hydrogenation) can controllably open a band gap in Si and Ge
by changing the orbital hybridization
\cite{Houssa98,Zheng7,Gao14,Ding100}, and hydrogenated germanene
(HGe) with a band gap of 1.53 eV has been synthesized
\cite{Bianco7}. The supporting substrates also have similar
functionalization effects on Si and Ge
\cite{Lin110,Chen110,Acun103,Cahangirov88,Yuan58}. These fast
synthesis and fabrication progresses lead to an urgent demand for
the knowledge of the thermal expansion and thermomechanics of
pristine/functionalized Si and Ge, because the accumulated thermal
strain and stress may influence the performance and lifetime of
devices working at finite temperatures.

\par In this letter, the phonon spectra, Gr\"uneisen constants, thermal expansion, and temperature-dependent stiffness of Si and Ge
are investigated by first-principles simulation. The roles of low dimensionality, layer thickness, and bond strength are
disentangled by analyzing the effects of chemical functionalization. The revealed physical picture is useful for understanding the
lattice dynamics of various two-dimensional systems.

The electronic structure and phonons are calculated using
density-functional methods \cite{methods}. The temperature dependent
lattice constant ($a$) is calculated using the self-consistent
quasiharmonic approximation (SC-QHA) method \cite{Huang2013}
\begin{equation}{\label{Equ_a_T}}
a(T)=\left(\frac{dE^e(a)}{da}\right)^{-1}\frac{1}{N_k}\sum_{k,\lambda}U_{k,\lambda}(T)\gamma_{k,\lambda},
\end{equation}
where $k$ and $\lambda$ are the wavevector and branch indices for a phonon mode; $N_k$ is the $k$-point number; $E^e$, $U_{k,\lambda}$, and $\gamma_{k,\lambda}$ are the total electronic energy, and
the internal energy and Gr\"uneisen constant
($\gamma=-\frac{a}{\omega}\frac{d\omega}{da}$) of a phonon mode, respectively. To
guarantee that $\gamma$ is calculated between two modes with the
same symmetry, the phononic $k{\cdot}p$ theory \cite{Huang2014} is
used to sort the phonon branches. Eq. (\ref{Equ_a_T}) is
self-consistently solved, and the phonon frequencies are updated
after each iteration step
($\Delta\omega\doteq-\gamma\frac{\Delta{a}}{a}\omega$). The
thermal-expansion coefficient ($\alpha=\frac{1}{a}\frac{da}{dT}$) can be re-expressed as \cite{Huang2013}
\begin{equation}{\label{Equ_expan_coeff}}
\alpha(T)=\frac{1}{\sum_{i=1}^3s_i}\frac{1}{N_k}\sum_{k,\lambda}C_V^{k,\lambda}(T)\gamma_{k,\lambda}\approx
\frac{2\sqrt{3}a^2}{B_{2D}N_k}\sum_{k,\lambda}C_V^{k,\lambda}(T)\gamma_{k,\lambda},
\end{equation}
where $C_V$ is the isovolume heat capacity and $B_{2D}$ is the two-dimensional bulk modulus
\begin{equation}{\label{Equ_bulk_modulus}}
B_{2D}(T)=A\frac{\partial^2F_{tot}}{\partial{A}^2}=\frac{1}{2\sqrt{3}a^2}\left[s_1-s_2+s_3+s_4\right],
\end{equation}
where $A$ ($=\frac{\sqrt{3}}{2}a^2$) is the unit area and
\begin{eqnarray}{\label{Equ_s1_3}}
s_1=a^2\frac{d^2E^e(a)}{da^2}, s_2=a\frac{dE^e(a)}{da},\nonumber \\
s_3=\frac{1}{N_k}\sum_{k,\lambda}\left[U_{k,\lambda}-TC_V^{k,\lambda}\right]\gamma^2_{k,\lambda},
s_4=\frac{2}{N_k}\sum_{k,\lambda}U_{k,\lambda}\gamma_{k,\lambda}.
\end{eqnarray}
The contributions of thermal expansion and phonon excitation to
$B_{2D}$ are included in ($s_1-s_2$) and ($s_3+s_4$), respectively,
and the phonon excitation is omitted in the quasistatic bulk modulus
($B^*_{2D}$) \cite{Wang2010}. Temperatures higher than 600 K are not
considered here, due to the low thermodynamic/kinetic stability of
these systems found in experiments
\cite{Enriquez24,Vogt108,Bianco7}, as well as to the appearance of
high-order anharmonicity that not included in QHA
\cite{Walle2002,Zakharchenko2009}.

\par
\begin{figure}[ht]
\scalebox{1.17}[1.17]{\includegraphics{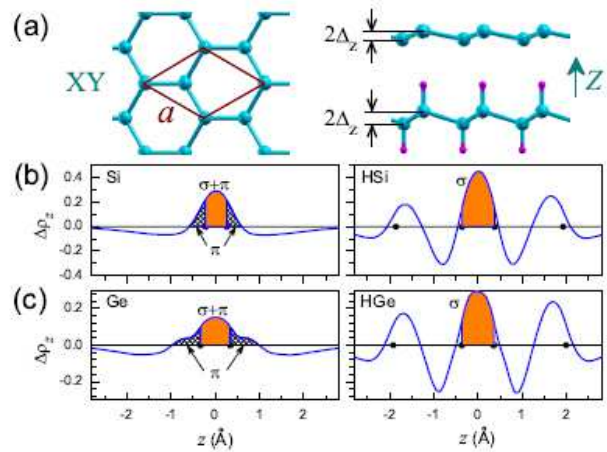}}
\caption{\label{Fig_struct_Q} (a) Structures and (b, c)
$\Delta\rho_z$ of pristine/hydrogenated Si and Ge. The origin
($\sigma$ and/or $\pi$ bondings) of the electron accumulation
($\Delta\rho_z>0$) near the Si/Ge atoms are indicated in shades.}
\end{figure}

 Hydrogenated Si and Ge (HSi and HGe) are used to study the effects of chemical functionalization. The structures are shown in
Fig. \ref{Fig_struct_Q}(a), and the calculated equilibrium lattice constants
($a$), buckling heights ($\Delta_z$), and bond lengths ($d$) are
listed in Tab. \ref{Tab_Struct}, which are consistent with other
experimental \cite{Enriquez24,Vogt108,Fleurence108,Bianco7} and
theoretical results \cite{Cahangirov102,Sahin80,Houssa98,Zheng7}.
The differential electron densities between the pre- and
post-bonding states ($\Delta{\rho}$) are projected onto the $z$
direction ($\Delta{\rho}_z$), which are shown in Fig.
\ref{Fig_struct_Q}(b) and (c). The $\sigma$ and $\pi$ bonds coexist
and compete with each other in buckled Si and Ge due to the
partial $sp^3$ orbital hybridization \cite{Cahangirov102,Sahin80}.
Therefore, apart from the space between the Si/Ge atoms
($|z|<\Delta_z$), the electrons also have some observable
accumulation ($\Delta\rho_z>0$) outside ($|z|>\Delta_z$) due to the
$\pi$ bonding (Fig. \ref{Fig_struct_Q}(b, c), left panels). Si has a
smaller $\Delta_z$ (0.22 \AA{}) than Ge (0.34 \AA{}), indicating the
less $sp^3$ hybridization in the former. Hydrogenation heightens
$\Delta_z$ (Tab. \ref{Tab_Struct}) and removes the $\pi$ electrons
(Fig. \ref{Fig_struct_Q}(b, c), right panels), due to the increased
$sp^3$ hybridization. The bond energy ($\epsilon_b$) is defined to
be the energy cost to break a Si--Si/Ge--Ge bond, which is used to
indicate the bond strength. Although the $\sigma$ bond is
strengthened by hydrogenation, the total strength ($\epsilon_b$) of
a Si--Si/Ge--Ge bond is decreased (Tab. \ref{Tab_Struct}), due to
the removal of the $\pi$ bond. On the other hand, hydrogenation
increases the dynamical stability of the bond, due to the removal of
the $\sigma$--$\pi$ competition. These changes in the bond
characteristics (hybridization, geometry, strength, and dynamical
stability) by hydrogenation will have profound effects on the
lattice dynamics.

\par
\begin{figure*}[ht]
\scalebox{1.16}[1.16]{\includegraphics{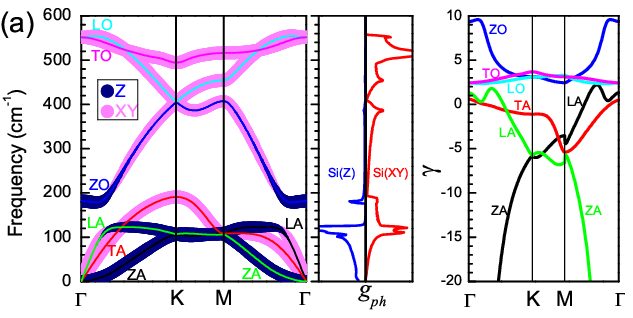}}
\scalebox{1.16}[1.16]{\includegraphics{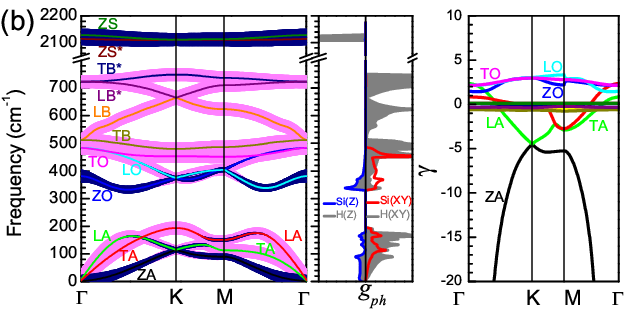}}
\scalebox{1.16}[1.16]{\includegraphics{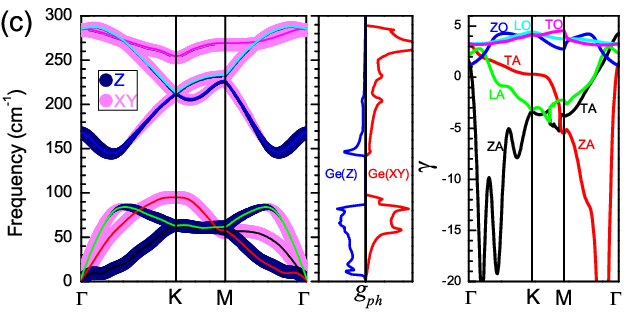}}
\scalebox{1.16}[1.16]{\includegraphics{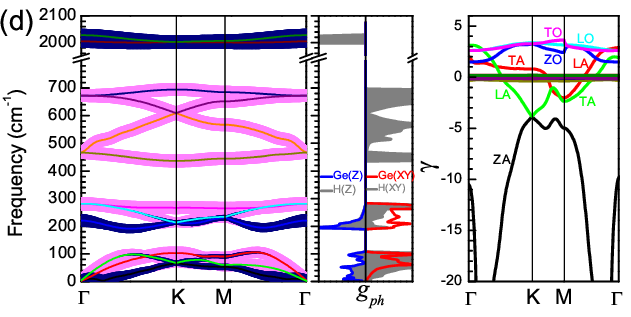}}
\caption{\label{Fig_phonons} Phonon dispersions, $g_{ph}$, and
$\gamma$ dispersions of (a) Si, (b) HSi, (c) Ge, and (d) HGe. The
line widths in the phonon dispersions are scaled by the amplitudes
of XY and Z vibrations, and the antisymmetric H modes in (b) and (d)
are labeled with the superscript "*".}
\end{figure*}

\begin{figure}[ht]
\scalebox{1.2}[1.2]{\includegraphics{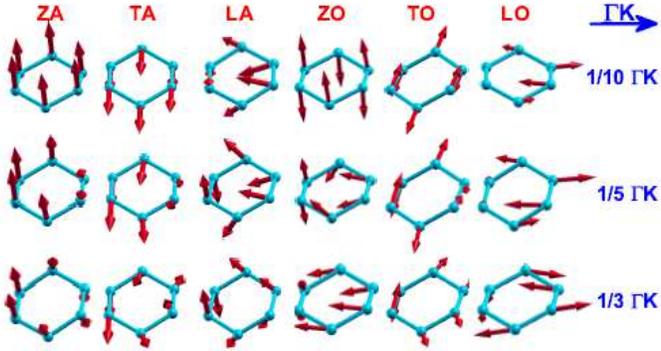}}
\caption{\label{Fig_vibrational_modes} The phonon modes in Si at
different $k$ points.}
\end{figure}

 A detailed knowledge on the vibrational states is a prerequisite
for understanding various phononic and thermodynamic properties. The
phonon dispersions and density of states ($g_{ph}$) of Si, HSi, Ge,
and HGe are shown in the left and middle panels of Fig.
\ref{Fig_phonons}, where the phonon branches are labeled according
to their initial symmetries at the $\Gamma$ point, and the
amplitudes of the in-plane (XY) and out-of-plane (Z) vibrations are
indicated by the line widths. The evolution of the phononic
eigenvectors of Si along the $\Gamma$K path are visualized in Fig.
\ref{Fig_vibrational_modes}.
\begin{table}[ht]
\caption{\label{Tab_Struct} Equilibrium lattice constant ($a$),
zero-point expansion ($\delta_{zp}$), buckling hight ($\Delta_z$), bond
lengths ($d_{\text{X--X}}$ and $d_{\text{X--H}}$, $\text{X}=$Si or
Ge), and bond energy ($\epsilon_b$) of Si, HSi, Ge, and HGe.}
\begin{ruledtabular}
\begin{tabular}{ccccccc}
    & $a$ (\AA{})& $\delta_{zp}$ (\%) & $\Delta_z$ (\AA{}) & $d_{\text{X--X}}$ (\AA{}) & $d_{\text{X--H}}$ (\AA{}) & $\epsilon_b$ (eV)\\\hline
Si  & 3.87       & 0.13               & 0.224              & 2.28                      & -                         & 2.91     \\
HSi & 3.89       & 0.11               & 0.359              & 2.36                      & 1.501                     & 2.51     \\
Ge  & 4.04       & 0.12               & 0.340              & 2.43                      & -                         & 2.41     \\
HGe & 4.07       & 0.13               & 0.367              & 2.46                      & 1.552                     & 2.00     \\
\end{tabular}
\end{ruledtabular}
\end{table}
The TA, TO, and LO branches in Si/Ge only consist of the XY
vibration, and the ZA branch of the Z vibration (Fig.
\ref{Fig_phonons}(a, c)). Thus, the atomic displacements in these
four modes (Fig. \ref{Fig_vibrational_modes}) are only modulated by
the Bloch phase factor ($e^{-i\bf{k\cdot{r}}}$). However, in the LA
and ZO branches, the XY and Z vibrations hybridize and swap with
each other at $k\sim0.22\Gamma$K/$\Gamma$M in Si
($0.40\Gamma$K/$\Gamma$M in Ge). This vibrational hybridization is
caused by the nonorthogonal covalent bonds, which also has been
observed in functionalized graphene \cite{Huang2014_2} and MoS$_2$
\cite{Huang2014} with nonorthogonal C--C and Mo--S bonds,
respectively. In planar graphene, the XY modes (TA, LA, TO, and LO)
are completely decoupled with the Z modes (ZA and ZO)
\cite{Huang2014_2}, because the C--C bonds in the XY plane are
orthogonal to the Z direction. The LA-ZO vibrational hybridization
changes the LA mode from XY to Z symmetry, which renders the LA
branch able to couple with the ZA branch at the Brillouin zone
boundary (Fig. \ref{Fig_phonons}(a, c)), where they are similar in
both frequency and symmetry. The LA-ZO and LA-ZA couplings result in
the flat dispersion of the LA branch at 110 cm$^{-1}$ in Si (60
cm$^{-1}$ in Ge). The acoustic/optical modes with Z vibration have
lower frequencies than their counterparts with XY vibration, because
the former ones mainly consist of the out-of-plane bending of the
Si--Si/Ge--Ge bonds, which are softer than the later ones consisting
of the in-plane bond bending/stretching. This is a low-dimensional
effect in phonon spectra. Ge has a larger buckling height $\Delta_z$
than Si, and the contribution of bond stretching in the ZO branch is
larger in more buckled layer, which tends to stiffen the ZO modes.
This is a thickness effect in the lattice dynamics of
low-dimensional systems \cite{Huang2013}.
Therefore, the LA-ZO coupling in Ge is smaller than that in Si,
resulting in less XY (more Z) vibration in the LA (ZO) branch and
larger dispersive LA branch in Ge (Fig. \ref{Fig_phonons}(c)).
Hydrogenation additionally introduces four bending (TB, TB$^*$, LB,
and LB$^*$) and two stretching modes (ZS and ZS$^*$), and the H
atoms also have some resonating displacements in the acoustic and
optical modes (Fig. \ref{Fig_phonons}(b, d)). Hydrogenation
increases the layer thickness ($\Delta_z$, Tab. \ref{Tab_Struct}),
which stiffens the ZO modes and enlarges the LA-ZO frequency gap.
The
$\frac{\omega_{\text{ZO}}-\omega_{\text{LA}}}{(\omega_{\text{ZO}}+\omega_{\text{LA}})/2}$
ratio is increased from 0.51 (0.51) in Si (Ge) up to 0.69 (0.65) in
HSi (HGe), which significantly decreases the LA-ZO coupling, and the
increases XY (Z) vibration in the LA (ZO) branch. With less Z
vibration in the LA branch after hydrogenation, the LA-ZA coupling
at the zone boundary is then decreased due to their enlarged
symmetry difference, making the LA branch there more dispersive. In
addition, the ZA branches in both Ge and HGe have wiggles near the
$\Gamma$ point, which indicates their low dynamical stability
(originating from weak $\sigma$ bond and $\sigma$-$\pi$
competition). This agrees with the facts that Ge still has not be
synthesized due to the preferred germanium-substrate alloying
\cite{Kara67}, and that HGe has a low amorphodization/decomposition
temperature ($\lesssim$348 K) \cite{Bianco7}. HGe is synthesized
prior to Ge, which may be due to the dynamical-stability enhancement
by hydrogenation.

\par The Gr\"uneisen-constant ($\gamma$) dispersions
are shown in the right panels of Fig. \ref{Fig_phonons}.
$\gamma$(ZA) has large negative values, which is ubiquitous for the
bending ZA modes in two-dimensional systems \cite{Huang2014}. These
bending ZA modes resemble the flexural vibration in guitar string,
where a tensile makes the lattice/string stiffer to such bending
vibration, resulting in a higher frequency and then a negative
$\gamma$. This is called guitar-string analogy \cite{Fultz2010} or
membrane effect \cite{Mounet2005}. Apart from $\gamma$(ZA),
$\gamma$(TA) and $\gamma$(LA) also have negative values in Si and Ge. In a TA
mode, the lattice is transversely distorted in the XY plane (Fig.
\ref{Fig_vibrational_modes}), and the restoring force comes from
both the in-plane bending and stretching of the covalent bonds. Bond
bending negatively contributes to $\gamma$ due to the guitar-string
analogy, while bond stretching has a normal positive contribution to
$\gamma$. When the bond is not strong enough, the effect of bond
stretching can not compete with that of bond bending,
resulting in negative net $\gamma$(TA). Diamond and graphene only
have positive $\gamma$(TA) \cite{Mounet2005,Huang2014_2} due to the
strong C--C bonds, while bulk silicon and germanium have negative
$\gamma$(TA) \cite{Xu1991,Wei1994} due to the relatively weak Si--Si
and Ge--Ge bonds. Through the LA-ZO and LA-ZA couplings, the weight
of Z vibration in the LA branch increases when leaving the $\Gamma$
point (Fig. \ref{Fig_phonons}(a, c), left panels), which increases
the contribution of out-of-plane bond bending, and then results in
negative $\gamma$(LA), especially at the zone boundary. All the
optical branches only have positive $\gamma$, due to the dominating
effect of bond stretching, which is similar in functionalized graphene \cite{Huang2014_2,Mounet2005}.
However, the $\gamma$(ZO) in planar graphene is negative, because of the dominating bond-bending
effect in those ZO modes. Except for the rippling part of the $\gamma$(ZA) dispersion in Ge, hydrogenation increases the $\gamma$
of ZA, TA, and LA modes in Si and Ge, because the layer thickening
increases the positive contribution of bond stretching in these
acoustic modes. The abnormal rippling part of the $\gamma$(ZA) in Ge
is caused by the low bond dynamical stability, and those ripples in
$\gamma$(ZA) are largely suppressed by hydrogenation, because the
related ZA modes become closer to normal two-dimensional bending modes. The
$\gamma$(ZO) of Si is exceptionally high ($=9$) near the $\Gamma$
point, due to the stretching of the nonorthogonal strong $\pi$ bonds
in the lowly-buckled Si lattice (Tab. \ref{Tab_Struct}).
Hydrogenation more or less decreases the $\gamma$ of ZO, TO, and LO
modes in Si and Ge, because the elimination of the $\pi$ bond
decreases the bond strength and then the effect of bond stretching.
The bending and stretching hydrogen modes (TB, TB$^*$, LB,
LB$^*$, ZS, and ZS$^*$) in HSi and HGe all have near-zero $\gamma$,
because the restoring forces on those perpendicular C--H bonds
are insensitive to the lateral-size variation. This means that
hydrogenation is an ideal approach to study the effect of chemical
functionalization on the lattice dynamics of Si and Ge, because
those modes specific to H atoms have negligible effects on the
lattice anharmonicity.

\par
\begin{figure}[ht]
\scalebox{0.83}[0.83]{\includegraphics{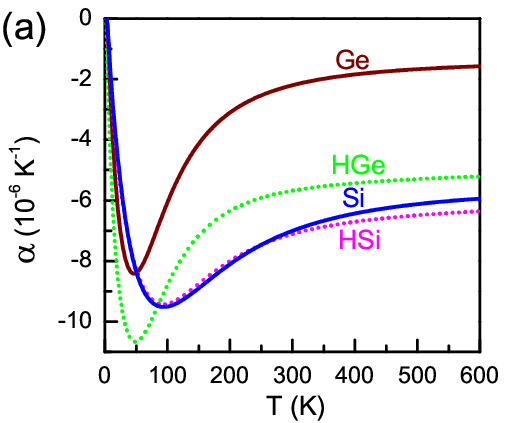}}
\scalebox{0.83}[0.83]{\includegraphics{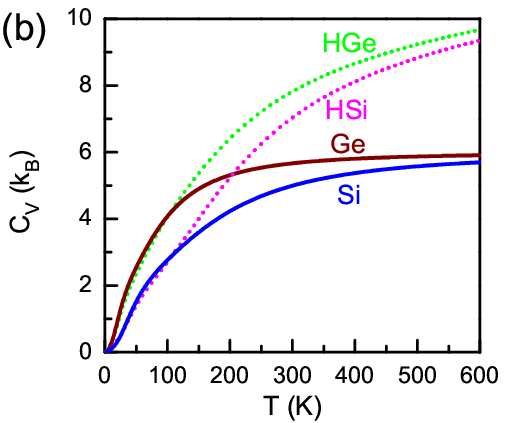}}\\
\scalebox{0.83}[0.83]{\includegraphics{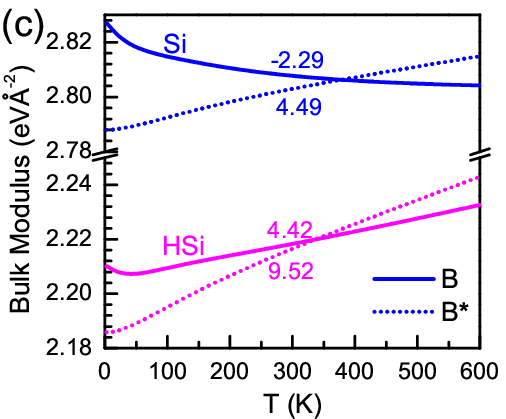}}
\scalebox{0.83}[0.83]{\includegraphics{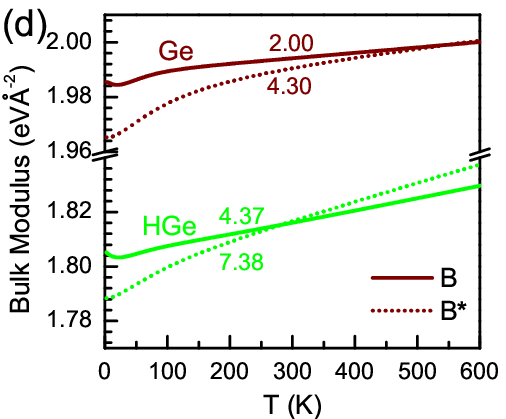}}
\caption{\label{Fig_dynamic_properties} Temperature dependence of
(a) $\alpha$, (b) $C_V$, and (c, d) $B_{2D}$ and $B_{2D}^*$ of Si,
HSi, Ge, and HGe. In (c) and (d), the curve slopes (in $10^{-5}$
eV\AA{}$^{-2}$K$^{-1}$) at 300 K are indicated.}
\end{figure}

 The above analyzed phononic properties can help understand various thermodynamic properties of Si and Ge, as
well as the effects of functionalization. The calculated
thermal-expansion coefficient ($\alpha$), isovolume heat capacity
($C_V$), and quasiharmonic/quasistatic bulk modulus
($B_{2D}$/$B_{2D}^*$) of Si, HSi, Ge, and HGe are shown in Fig.
\ref{Fig_dynamic_properties}. A fully excited phonon branch
contributes 1.0 $k_B$ to $C_V$, and the nominal number of excited
branches is about three at 100 (50) K in both Si and HSi (Ge and
HGe) (Fig. \ref{Fig_dynamic_properties}(b)). This means that the
excitation of the three acoustic branches (ZA, TA, and LA) dominates
at temperatures lower than 100 (50) K, and the optical modes (ZO,
TO, and LO) are only considerably excited at higher temperatures.
Acoustic modes mostly have negative $\gamma$, while optical modes
only have positive $\gamma$ (Fig. \ref{Fig_phonons}, right panels).
Due to the subsequent excitation of these negative- and
positive-$\gamma$ modes, $\alpha$ firstly decreases with increasing
temperature until 100 (50) K (Fig. \ref{Fig_dynamic_properties}(a)),
and then increases. However, $\alpha$ is still consistently
negative, due to the dominating contribution from the
negative-$\gamma$ modes. Although the excitation of the additional H
modes makes the $C_V$ of HSi (HGe) larger than that of Si (Ge) above
120 K (Fig. \ref{Fig_dynamic_properties}(b)), its effect on the
minimum-$\alpha$ temperature is negligible due to their near-zero
$\gamma$ of these modes (Fig. \ref{Fig_phonons}(b, d), right
panels), and hydrogenation also has a negligible effect on the
expansion caused by the zero-point vibrations ($\delta_{zp}$, Tab.
\ref{Tab_Struct}). Therefore, the hydrogenation effect on the
lattice anharmonicity only comes from the chemical functionalization
on the Si--Si/Ge--Ge bond. Although the hydrogenation-induced
decreases of $B_{2D}$ (by 30\%, Fig.
\ref{Fig_dynamic_properties}(c)) and optical $\gamma$ (Fig.
\ref{Fig_phonons}(a, b)) tend to further decrease the negative
$\alpha$ (Eq. \ref{Equ_expan_coeff}), $\alpha$(HSi) is still close
to $\alpha$(Si),
due to the canceling effect from the increase of acoustic $\gamma$
(Fig. \ref{Fig_phonons}(a, b)). $\alpha$(Ge) is much higher than
$\alpha$(HGe) (by $4\times10^{-6}$K$^{-1}$ above 300 K), because
$\gamma$(ZA) is significantly decreased by hydrogenation (Fig.
\ref{Fig_phonons}(c, d)), which negatively contributes to $\alpha$.

\par Thermal expansion is caused by the excitation of anharmonic phonons, while both thermal expansion and phononic excitation have individual
contributions to the thermomechanics (Eq. \ref{Equ_bulk_modulus} and
\ref{Equ_s1_3}). Both contributions are considered in the
temperature dependence of $B_{2D}$, while only the former
in that of $B_{2D}^*$. The negative thermal expansion results in the
increase of $B_{2D}^*$ under heating, due to the increased curvature
of the potential-energy surface under lattice contraction (i.e.,
$\frac{dB_{2D}^*}{da}<0$). The stiffening effect from the zero-point
vibrations makes $B_{2D}$ larger than $B_{2D}^*$ at low
temperatures. However, $B_{2D}$(T) curves have lower increasing
rates than their $B_{2D}^*$(T) counterparts, and the $B_{2D}$
of Si even decreases under heating, which is due to the softening
effect from the excitation of negative-$\gamma$ modes ($s_4$ term in
Eq. \ref{Equ_bulk_modulus}). Therefore, thermal expansion and phonon
excitation have reverse effects on $B_{2D}$, which also has been
observed in positively expanding MoS$_2$ \cite{Huang2014}. Although
hydrogenation decreases $B_{2D}$ and $B_{2D}^*$ due to the bond
weakening (Tab. \ref{Tab_Struct}), hydrogenation increases both the
slopes of $B_{2D}$(T) and $B_{2D}^*$(T) curves
($\frac{dB_{2D}}{dT}$ and $\frac{dB^*_{2D}}{dT}$).
$\frac{dB_{2D}^*}{dT}$ equals $a\frac{dB_{2D}^*}{da}\times\alpha$,
and the $\frac{dB^*_{2D}}{dT}$ of Si at 300 K is increased by
5.0$\times10^{-5}$ eV\AA{}$^{-2}$K$^{-1}$. As Si and HSi have the same $\alpha$, this increase is only ascribed to the
enlargement of potential-surface anharmonicity, namely, to the
decrease in $a\frac{dB_{2D}^*}{da}$ (from -6.3 to -13.3
eV\AA{}$^{-2}$). The increase in the $\frac{dB_{2D}}{dT}$ of Si by hydrogenation is even
larger (by 6.7$\times10^{-5}$ eV\AA{}$^{-2}$K$^{-1}$) than
$\frac{dB^*_{2D}}{dT}$, due to the increased acoustic $\gamma$ ($\frac{dB_{2D}}{dT}\sim\sum{C_V^{k,\lambda}\gamma_{k,\lambda}}$). The decrease in
the negative $\alpha$ of Ge by hydrogenation contributes to the increase of
$\frac{dB^*_{2D}}{dT}$ (by 3.1$\times10^{-5}$
eV\AA{}$^{-2}$K$^{-1}$ at 300 K), which is partially canceled by the
increase in $a\frac{dB_{2D}^*}{da}$ (from -18.6 to -13.8
eV\AA{}$^{-2}$). The increase in $\frac{dB_{2D}}{dT}$ (by
2.4$\times10^{-5}$ eV\AA{}$^{-2}$K$^{-1}$) is smaller than that of
$\frac{dB^*_{2D}}{dT}$, due to the decrease in $\gamma$(ZA) after
the dynamical-stability enhancement by hydrogenation (Fig.
\ref{Fig_phonons}(c, d)).

In summary, the roles of lattice dimensionality and bond
characteristics in the lattice dynamics of
silicene and germanene, as well as in the chemical functionalization
effects, have been revealed. For the substrate-supported silicene and germanene \cite{Lin110,Chen110,Acun103,Cahangirov88,Yuan58},
as well as other two-dimensional materials, the environmental interactions may
bring some complicated structural and electronic influences \cite{Gao592,Liu117,Cai2013}. The physical picture and analysis methods established here for
the correlation between structure, electronics, and lattice dynamics will be useful for studying the corresponding lattice anharmonicity therein.

This work is supported by the National Science Foundation of China
under Grant No. 11204305 and U1230202(NSAF), and the special Funds
for Major State Basic Research Project of China (973) under Grant
No. 2012CB933702. The calculations were performed in Center for
Computational Science of CASHIPS and on the ScGrid of Supercomputing
Center, Computer Network Information Center of CAS.

\bibliographystyle{apsrev}
\bibliography{Reference_list}

\end{document}